\documentclass[twocolumn,tighten,times,useAMS,usenatbib]{aastex631}

\usepackage{epstopdf,soul}

\graphicspath{{./}{figures/}}
\shorttitle{High-resolution spectroscopy of SN~2023ixf}
\shortauthors{Smith et al.}

\begin{document}
\title{High resolution spectroscopy of SN~2023ixf's first week: Engulfing the Asymmetric Circumstellar Material}

\author[0000-0001-5510-2424]{Nathan Smith}
\affil{Steward Observatory, University of Arizona, 933 N. Cherry Avenue, Tucson, AZ 85721, USA}

\author[0000-0002-0744-0047]{Jeniveve Pearson}
\affil{Steward Observatory, University of Arizona, 933 N. Cherry Avenue, Tucson, AZ 85721, USA}

\author[0000-0003-4102-380X]{David J.\ Sand}
\affil{Steward Observatory, University of Arizona, 933 N. Cherry Avenue, Tucson, AZ 85721, USA}

\author[0000-0002-0551-046X]{Ilya Ilyin} \affiliation{Leibniz-Institut
  f\"ur Astrophysik Potsdam (AIP), An der Sternwarte 16, D-14482
  Potsdam, Germany}

\author[0000-0002-4924-444X]{K.\ Azalee Bostroem}
\affil{Steward Observatory, University of Arizona, 933 N. Cherry Avenue, Tucson, AZ 85721, USA}
\altaffiliation{LSSTC Catalyst Fellow}

\author[0000-0002-0832-2974]{Griffin Hosseinzadeh}
\affil{Steward Observatory, University of Arizona, 933 N. Cherry Avenue, Tucson, AZ 85721, USA}

\author[0000-0002-4022-1874]{Manisha Shrestha}
\affil{Steward Observatory, University of Arizona, 933 N. Cherry Avenue, Tucson, AZ 85721, USA}


\begin{abstract}

We present a series of high-resolution echelle
spectra of SN~2023ixf in M101, obtained nightly during the
first week or so after discovery using PEPSI on the LBT.
Na~{\sc i}~D absorption in these spectra indicates a host reddening of $E(B-V)$=0.031~mag
and a systemic velocity of $+$7~km~s$^{-1}$ relative to the
average redshift of M101.  Dramatic changes are seen in
in the strength and shape of strong emission lines
emitted by circumstellar material~(CSM), including He~{\sc ii}~$\lambda$4686, 
C~{\sc iv}~$\lambda\lambda$5801,5811, H$\alpha$, and
N~{\sc iv}~$\lambda\lambda$7109,7123.  In general, these narrow lines
broaden to become intermediate-width lines before disappearing from
the spectrum within a few days, indicating a limited extent to
the dense CSM of around 20-30 AU (or $\la$10$^{14.7}$ cm).  H$\alpha$ persists in the spectrum
for about a week as an intermediate-width emission line with P~Cyg absorption at 700-1300 km
s$^{-1}$ arising in the post-shock shell of swept-up CSM.  Early
narrow emission lines are blueshifted and indicate an expansion speed
in the pre-shock CSM of about 115 km s$^{-1}$, but with even broader
emission in higher ionization lines.  This is faster than the
normal winds of red supergiants, suggesting some mode
of eruptive mass loss from the progenitor or radiative
acceleration of the CSM.  A lack of narrow blueshifted absorption
suggests that most of the CSM is not along our line
of sight.  This and several other clues indicate that the CSM of
SN~2023ixf is significantly aspherical.  We find that CSM lines disappear after a few days because the 
asymmetric CSM is engulfed by the SN photosphere.

\end{abstract}

\keywords{supernovae: individual (SN~2023ixf) --- circumstellar material}

\section{Introduction} \label{sec:intro}

Understanding the late evolution and end fates of massive stars
remains an enduring challenge.  It was recognized long ago that mass
loss plays a key role in determining the outcome of stellar evolution
\citep{p71,cm86}.  In recent years, however, the traditional view
where well-behaved, steady stellar winds of single stars lead to
predictable outcomes with reliable metallicity-dependence
\citep[e.g.,][]{heger03}, has gradually been eroding, giving way
instead to a more complicated picture where binary interaction and
eruptive events dominate the mass loss \citep[see][for a
  review]{smith14araa}.  These modes of mass loss do not have
well-established monotonic trends with initial mass or metallicity,
and are challenging for models of
single-star evolution.

A major reason for this shifting paradigm is that normal, steady
stellar winds of hot massive stars are evidently not as strong as we
used to think, reducing their ability to remove the H envelope and to
strongly impact evolution \citep{so06,puls08,smith14araa,sundqvist19}. The
same applies to RSG winds, indicated by recent downward revisions of
normal RSG wind mass-loss rates and the general scarcity of
dust-enshrouded RSGs \citep{beasor20,bs22}.  This shift is also
influenced by results from several different lines of inquiry: (1)
strong evidence for binary-stripped progenitors of H-poor supernovae
(SNe) \citep{ppod92,smith11,drout11}, (2) observational evidence for
extreme, eruptive modes of mass loss \citep{so06}, and (3) firmer
observational estimates of a high interacting binary fraction among O-type
stars \citep{sana12}.

Another driving factor toward a more complicated view of mass loss has
been the discovery of a number of different explosive transients that
simply do not fit predictions of the traditional view of a massive star
evolution dominated by single-star wind mass loss.  Chief among these
are SNe with signatures of strong shock interaction with circumstellar
material (CSM).  Evolved massive stars that retain their H envelopes
usually have large radii and relatively slow escape speeds, which can
lead to slow CSM that produces narrow H lines in the spectrum of the
SN.  In this case, they are classified as Type~IIn
\citep{schlegel90}. The illumination or shock heating of close-in CSM can provide
unique clues about the mass-loss properties of the progenitor star in
the late evolutionary phases of its life, which are otherwise
difficult to infer \citep[see][for a review]{smith17review}.  There is 
wide diversity among SNe with observed signatures of
H-rich CSM interaction, ranging from super-luminous SNe IIn, scaling 
down through normal SNe IIn, and further down to those with barely 
any observable signatures of CSM interaction.  

On the less extreme end, we see interacting SNe where the spectral
signatures of CSM interaction are fleeting.  One of these events is the topic
of the current paper.  The narrow lines may last for
only a few days or a week before fading, and they can quickly
transition to look like normal\footnote{``Normal'' here means a
  visual-wavelength spectrum dominated by an ejecta photosphere, not
  by CSM interaction.} SNe.  This class of objects has been known for
about four decades.  Well-studied examples of the phenomenon include
SN~1983K \citep{niemala85}, SN~1993J \citep{benetti94,gh94}, SN~1998S
\citep{shivvers15}, SN~2006bp \citep{quimby07}, PTF11iqb
\citep{smith11iqb}, and SN~2013cu \citep{galyam14,groh14,gv16}.
Additional events studied in detail include SN~2013fs, SN~2017ahn, SN 2020pni, and SN~2020tlf \citep{bullivant18,tart21,terreran22,wynn22}.  Because study of this class requires early discovery on timescales of
hours or days after explosion, early examples were limited to
fortuitous early detections of nearby events.  With more systematic
transient searches and early discovery becoming more routine, growing
samples of this class have been identified \citep{khazov16}.  Some
estimates suggest that a large fraction ($\gtrsim1/3$) of
otherwise normal core-collapse SNe (ccSNe) have these early CSM features \citep{bruch21},
which is larger than the 8-9\% of ccSNe that are more
traditional strongly-interacting SNe~IIn \citep{smith11}.

The defining characteristic of this class is very short-lived (a few
days) narrow emission lines in the spectrum, which are thought to
result from dense and confined CSM within 10s of AU around the
progenitor star.  As with the broader class of SNe IIn, this is a
phenomenon that is not unique to any one type of explosion or any
unique progenitor type because it depends on the characteristics of the
surrounding material --- in principle, any SN type might be surrounded
by dense and confined CSM.  In practice, the observed events tend to
be H rich (perhaps because narrow lines require that a progenitor had
slow escape speeds due to a large H envelope), and they usually evolve
into SNe IIb, II-P, or II-L when the CSM interaction signatures fade.
Early spectra show high-ionization emission lines like He~{\sc ii} and
doubly or triply ionized C and N lines with narrow cores and broad
wings, and these emission lines sit atop a smooth blue continuum.  The
high ionization level is thought to arise from photoionization of the
CSM by a hard radiation field, produced either by a UV/X-ray flash
from shock breakout, or produced by the shock when the fastest SN
ejecta first crash into the CSM.  These high-ionization lines cause
the early spectra to resemble Wolf-Rayet (WR) stars, leading to some
claims that this points to WR progenitors \citep{galyam14}.  However,
the WR spectral features arise because a slow, dense, H-rich wind is
ionized by the SN; the progenitor star is likely to have been cool and
potentially even self-obscured by its CSM, and would not have been
seen as a WR star, more likely resembling a cool hypergiant
\citep{smith11iqb,groh14}.  Of course, this would depend on when
exactly the progenitor star was observed, since the confined CSM may
have just been produced shortly before the SN (i.e. the star may have
appeared as a normal RSG or YSG a few years earlier).

There are several remaining open questions about this class of
objects, concerning the mechanism that ionized the CSM (flash from
shock breakout or shock interaction), the range of physical properties of
the CSM (total CSM mass or mass-loss rate of the progenitor, range of
shell/envelope radii, asymmetry, composition, etc.), timescale of the
mass-loss before explosion, details of the evolution of the shock
through the CSM, range of initial masses for the progenitors, and so
on.  All of these help to inform the most important question, which
concerns the physical mechanism operating within the star that caused
it to suddenly eject so much mass right before core collapse.  The
observed velocities of the CSM and the quick disappearance of the
narrow lines (and hence, the small inferred outer boundary of the CSM)
imply that the strong mass loss occurred very soon before core
collapse, perhaps in the last few months or the final year or two of
the star's life.
This timescale is a strong hint that something is going haywire in the
star during the last rapid phases of nuclear burning (Ne, O, or Si
burning), and several ideas have been proposed for extreme mass loss
triggered during these phases
\citep{arnett_turbulent_2011,quataert_wave-driven_2012,shiode_observational_2013,shiode_setting_2014,sa14,woosley_remarkable_2015,fuller_pre-supernova_2017,wu_diversity_2021}.
Since the CSM interaction is so short-lived (and the total CSM mass
estimates are on the order of 0.1 $M_{\odot}$), and as these objects
evolve into relatively normal SN types when the narrow lines fade
(perhaps implicating moderately massive 10-20 $M_{\odot}$ red or
yellow supergiant progenitors), it is unlikely that some other
mechanisms proposed for pre-SN mass loss in SNe~IIn will be applicable
to this particular class.  For instance, pulsational-pair instability
eruptions \citep{woosley17} are limited to only very high initial
masses, and are probably ruled out for these objects.  Also, it is
difficult to understand why strong pulsationally driven superwinds
from very luminous RSGs \citep{yoon_evolution_2010} would only operate
for $\sim$1 yr before core collapse.
In any case, the range of physical parameters for the CSM deduced from
studies of individual events can help inform what mechanism ejected
the CSM. Perhaps it can also help to understand how/if these objects
are connected to the broader class of interacting SNe IIn, or if they
are a distinct phenomenon.

\begin{deluxetable}{ l c c c}
\centering
\caption{Log of LBT/PEPSI Observations} \label{tab:PEPSIlog}
\tablehead{\colhead{Date (UTC)} & \colhead{MJD} & \colhead{Epoch (days)} & \colhead{Airmass}}
\startdata
2023-05-21 & 60085.373  & 2.62  & 1.35  \\
2023-05-22 & 60086.244  & 3.49  & 1.08  \\
2023-05-23 & 60087.150  & 4.40  & 1.14  \\
2023-05-24 & 60088.155  & 5.40  & 1.12  \\
2023-05-26 & 60090.357  & 7.60  & 1.34  \\
2023-05-27 & 60091.183  & 8.43  & 1.08  \\
2023-06-05 & 60100.329  & 17.56 & 1.34  \\
\enddata
\end{deluxetable}

Here we discuss a new member of this class, SN~2023ixf, which exploded
in the very nearby spiral galaxy M101.  It was discovered by
K.\ Itagaki on 2023 May 19, and was soon classified as a Type II SN by
\citet{perley23}.  In the following, we adopt a 
host redshift for M101 of $z$=0.000804 \citep{devaucouleurs_vizier_1995}.
From examining pre-explosion archival images, a candidate progenitor
consistent with a moderate-luminosity RSG progenitor has been
identified \citep[][Jencson et al. in perp.]{ps23,soraisam23,kilpatrick_sn2023ixf_2023}, suggesting a star that had an
initial mass of around 12-17 $M_{\odot}$.

\begin{figure*}
    \centering
    \includegraphics[width=6.3in]{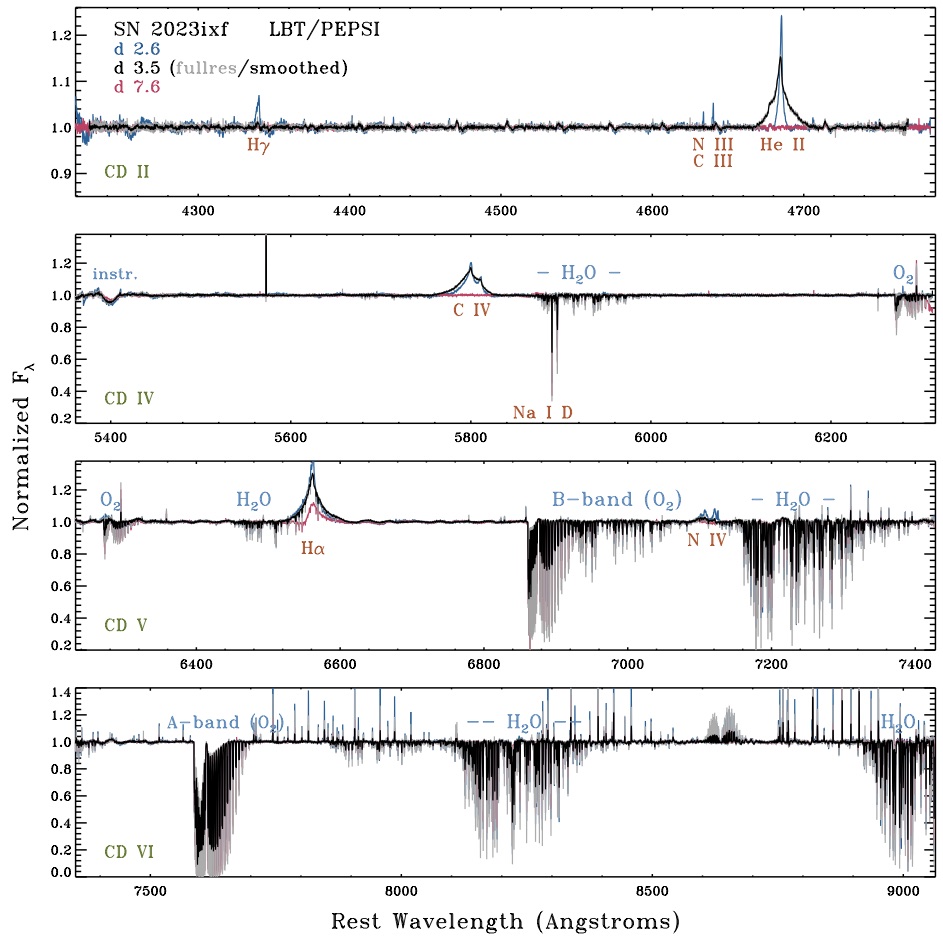}
   \caption{An example of the full wavelength coverage of the PEPSI
     spectra, showing the first epoch on day 2.6 in blue, the second epoch on day 3.5 in grey
     (original resolution) and black (smoothed). The
     spectrum a few days later on day 7.6 is shown in magenta,
     when the narrow features besides H$\alpha$ have faded.  Telluric
     absorption bands or instrumental features are labeled in blue, lines intrinsic to the SN
     (or host galaxy) are labeled in red-orange, and the PEPSI cross
     disperser (CD) is noted in green in the lower left of each panel.
     Each of the four wavelength ranges shown here results from
     stitching together several orders, sometimes producing small
     wiggles in the continuum at locations where edges of orders are
     joined.  This is especially noticeable in CD II with a pattern of
     wiggles every $\sim$30 \AA.  One of these is on the blue wing of
     He~{\sc ii} $\lambda$4686.  The dip in flux at 5400 \AA \, is a known instrumental artifact of PEPSI.}
   \label{fig:full}
\end{figure*}

SN~2023ixf was quickly rising at the time of discovery and was expected to become very
bright, and because it was a Type II event that could potentially show
early narrow lines in the spectra, we chose to initiate an intensive
observing campaign to obtain high-resolution echelle spectra every
night (or almost every night) for the first week or so after
discovery, in order to document rapid changes in the narrow emission
from CSM.  These observations and initial results are described here,
while companion papers describe the early light curve \citep{griffin23ixf}
and low-resolution spectra (Bostroem et al., in prep.).  Section~\ref{sec:obs}
describes the observations, Section~\ref{sec:res} describes the resulting
data and analysis, and Section~\ref{sec:disc} presents our
interpretation of these early data.

\section{Observations} \label{sec:obs}

Shortly after discovery, we initiated a campaign to obtain
observations of SN~2023ixf with a nearly nightly cadence using the
Potsdam Echelle Polarimetric and Spectroscopic Instrument (PEPSI;
\citealt{pepsi}) mounted on the Large Binocular Telescope (LBT)
located on Mt. Graham, AZ.  PEPSI is a cross-dispersed echelle
spectrograph with separate blue and red channels, each with three
wavelength ranges corresponding to three cross dispersers (CDs), with
CD I, II, and III in the blue arm, and CD IV, V, and VI in the red
arm.  When combined, these are designed to cover the full optical
wavelength range with no gaps.  At the time of these observations, CD
I and CD III were not available, so we used CD II (covering 4219-4787
\AA) in the blue channel, and CD IV (5361-6316 \AA), CD V (6232-7428
\AA) and CD VI (7351-9064 \AA) in the red channel. All observations were composed of a 60 minute blue channel exposure with CD II and three 20 minute red channel exposures with CD IV, CD V, and CD VI. We used a 300
$\mu$m fiber (2.2 arcsec diameter) corresponding to a spectral
resolving power of $R$=$\lambda/\Delta\lambda$=50,000, or a velocity
resolution of about 6 km s$^{-1}$.  

The data were reduced using the
Spectroscopic Data Systems (SDS) pipeline \citep{ilyin00,pepsi}.  
The pipeline performs bias subtraction and flat field correction, order
tracing and optimal extraction with cosmic ray elimination, and wavelength
calibration. The spectral orders are normalized in 2D with a non-linear
constrained least-squares fit to account for broad emission lines
spanned over adjacent spectral orders. Finally, the spectral orders
are rectified into a single spectrum for each CD. The wavelength scale
was reduced to the Solar System Barycentric rest frame using
JPL ephemerides.  The pipeline also estimates the variance in each pixel.

Based on early photometry and upper limits, \citep{griffin23ixf} estimate a likely explosion time of MJD=60082.75.
Using this as a reference, our LBT/PEPSI spectra were obtained between 2 and 9
days after explosion, plus one additional later epoch.  We use this date to calculate the time since 
explosion for each PEPSI spectrum, listed as the 3rd column in Table 1.

\begin{figure}
    \centering
    \includegraphics[width=3.0in]{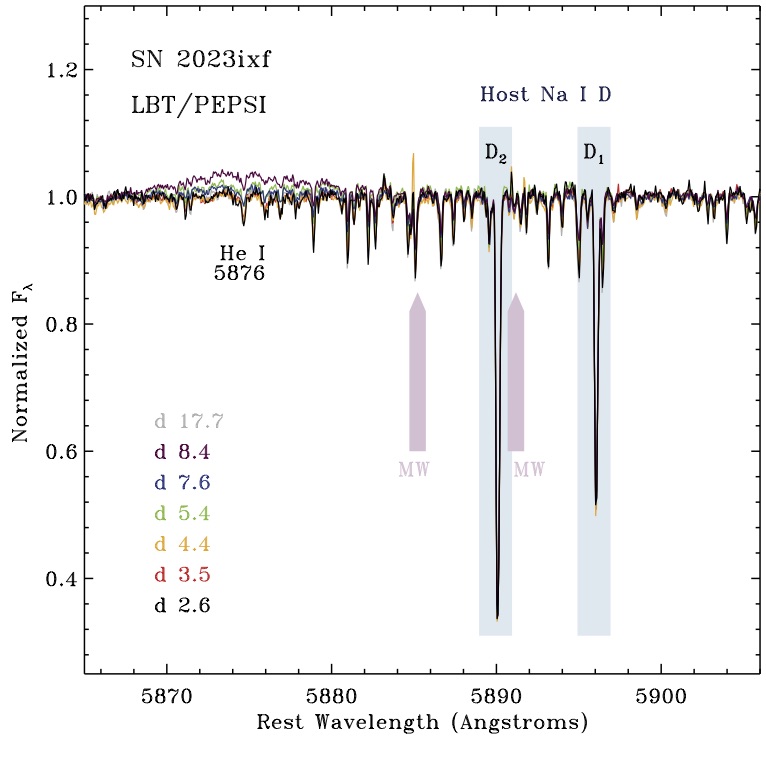}
   \caption{Detail of the region around the Na~{\sc i} D interstellar
     absorption doublet seen toward SN~2023ixf in PEPSI spectra. The light blue boxes show the expected locations of the Na I lines for the host, and the lavender pointed boxes show the Na I line locations for zero redshift, appropriate for features in the Milky Way (blueshifted here because the spectra have been corrected for the redshift of M101). 
     Changes in intermediate-width emission from He~{\sc i}
     $\lambda$5876 are also visible, and are discussed in more detail
     in Section \ref{sec:ion}. }
   \label{fig:naid}
\end{figure}

\section{Results} \label{sec:res}

An example of the resulting normalized PEPSI spectrum of SN~2023ixf is
shown in Figure~\ref{fig:full}.  This shows the spectra on days 2.6 (blue), 
3.5 (black), and 7.6 (magenta).  Although the spectrum appears somewhat complicated, most
of the structure results from complex telluric absorption bands, which
are labeled in blue in Figure~\ref{fig:full}.  Overall, the spectrum
of SN ~2023ixf at these early times is dominated by a very smooth blue
continuum (although the continuum slope is normalized here) plus a
small number of prominent lines labeled in red-orange in
Figure~\ref{fig:full}.  At this early epoch within only about a week after
explosion, it does not yet show any emission or absorption from very
broad features associated with the fast SN ejecta.  Besides the
interstellar absorption from Na~{\sc i} D, the most interesting
features are the strong narrow emission lines in the spectrum that
dramatically change strength and shape with time over only a few
days: He~{\sc ii} $\lambda$4686, C~{\sc iv} $\lambda\lambda$5801,5811,
and H$\alpha$.  The spectra also show weak narrow emission N~{\sc iii} $\lambda\lambda$4634,4641 and C~{\sc iii} $\lambda\lambda$4648,4650 seen only in our first epoch, as well as weak emission from N~{\sc iv}
$\lambda\lambda$7109,7123 and He~{\sc i} $\lambda$5876, the latter of
which grows in strength with time (see Fig.~\ref{fig:naid}).  We discuss each of these lines in turn, after
briefly examining the narrow interstellar absorption.

\begin{figure*}
    \centering
    \includegraphics[width=6.3in]{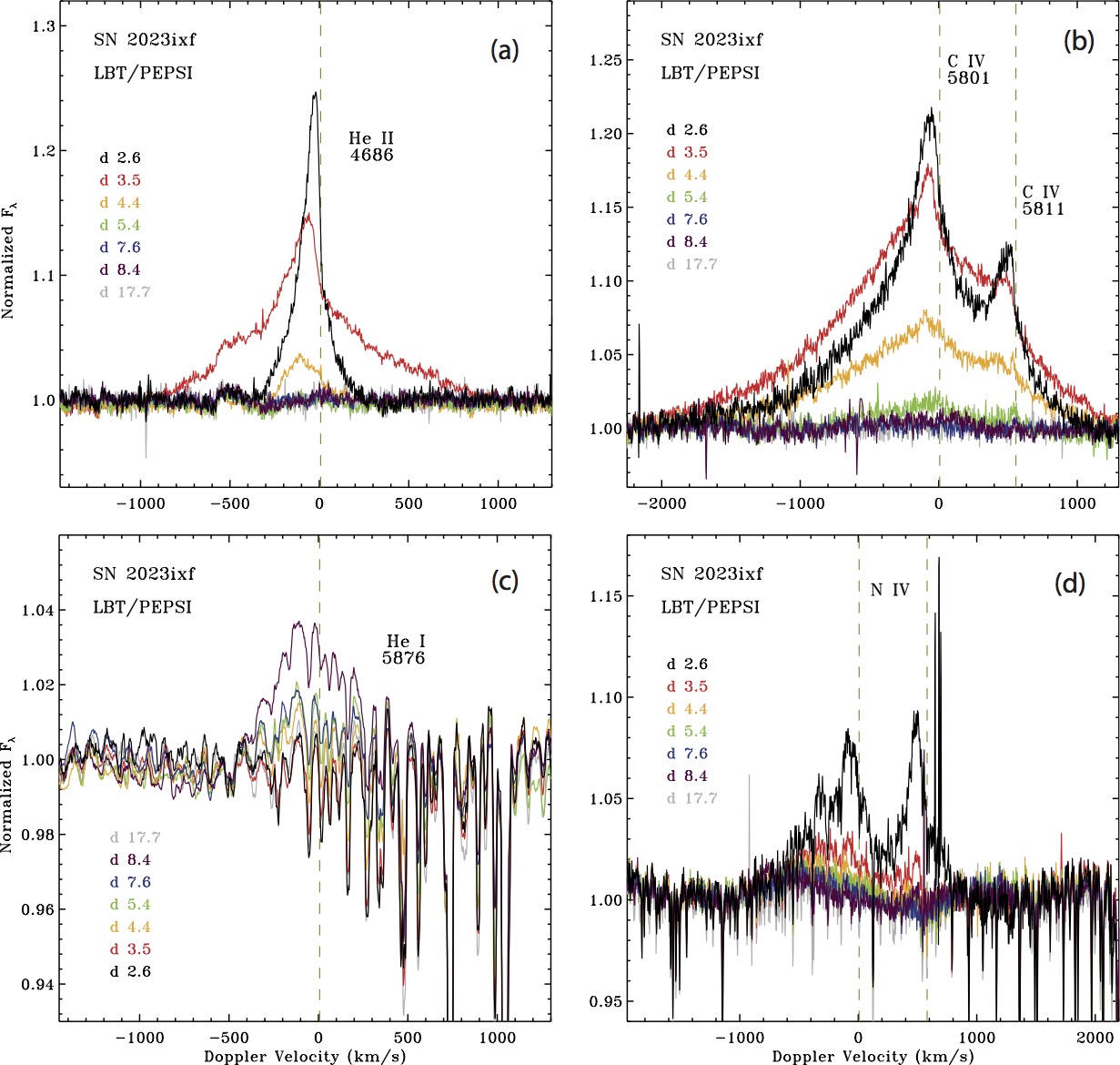}
   \caption{Evolution of individual line profiles for (a) He~{\sc ii}
     $\lambda$4686, (b) the C~{\sc iv} $\lambda\lambda$5801,5811
     doublet (velocity scale set for $\lambda$5801), (c) He~{\sc i}
     $\lambda$5876, and (d) the N~{\sc iv} $\lambda\lambda$7109,7123
     doublet (velocity scale set for $\lambda$7123).  Spectra have been 
     corrected for the redshift of M101, $z$=0.000804, and for each line,
     velocity = +7 km s$^{-1}$ (the presumed systemic velocity) is shown 
     by a vertical green dashed line (see text). Note that for the day 8.4 
     spectrum of C IV (panel b), there
     was no line flux detected, but there were a few noise spikes or
     hot pixels in the spectrum that we clipped at a normalized flux
     level of 1.02 for clarity in this figure.}
   \label{fig:quad}
\end{figure*}

\subsection{Na I D and Interstellar Reddening}

Because the line of sight Milky Way reddening toward M101 of
$E(B-V)$=0.0074 mag \citep{sf11ebv} is quite low and because we are
primarily examining line profiles in normalized spectra (where the
continuum flux is divided out anyway), we make no reddening correction
to our echelle spectra.  However, high-resolution echelle spectra afford an
opportunity to provide a precise constraint on the equivalent width
(EW) of the narrow Na~{\sc i} D interstellar absorption lines arising
along the line of sight through the host galaxy.

Figure~\ref{fig:naid} shows a detail of the region of the spectrum
around the Na~{\sc i} D$_1$ and D$_2$ resonance doublet, which has
been corrected for a redshift of $z$=0.000804.  Each epoch of PEPSI
spectra is shown, and the nature of narrow absorption lines in the
spectrum is remarkably consistent over the various epochs.  The blue
boxes in Figure~\ref{fig:naid} indicate the expected positions of the
two Na~{\sc i} lines in M101.  Indeed there are two strong narrow
absorption features detected here.  These are seen to be mixed amid a large
number of other weaker narrow absorption features, which may arise
from various clouds along the line of sight through the Milky Way at
various rotation velocities and a cluster of telluric water absorption lines.  This suggests that high spectral
resolution is important to accurately estimate the reddening within
M101, because its low redshift means that its interstellar
absorption features may overlap with those from the Milky Way and these telluric features.

We measure Na~{\sc i} EW values of 0.156 \AA  \, for D$_1$ and 0.187 \AA \,
for D$_2$.  The measured EWs were consistent to better than 1\% from
one epoch to the next; the main uncertainty in the absolute
measurement of the EW values is the choice of continuum level, and how
much contamination there might be from overlapping lines.  Although
there are many other weak narrow lines, the continuum level in these
spectra is well determined to about 1\% of the flux level as well.
Following the relations from \citet{dovi12}, these EWs translate to a
reddening value of $E(B-V)$ = 0.036 mag.  Following \citet{dovi12}, we
multiply this by 0.86 to account for the conversion to \citet{sf11ebv}
values, yielding a host reddening along the line of sight to
SN~2023ixf of $E(B-V)$ = 0.031 mag.  The uncertainty in the relation from
\citet{dovi12} is 30-40 \%, which is much larger than any error in
$E(B-V)$ introduced by measurement error in these PEPSI spectra.  This
resulting value of $E(B-V)$ = 0.031 mag is larger than the Milky Way
reddening, indicating that the local host dust in M101 is the dominant
source of extinction along the line of sight.  This may, of course,
vastly underestimate the extinction toward the progenitor star arising from dust that
may have been present in the pre-SN CSM.

The velocities of these deep Na~{\sc i} D absorption features are also
useful later in our analysis, especially when interpreting velocities
of narrow emission components.  We measure the centroid velocity of
the strongest components of the D$_1$ and D$_2$ lines.  After
correcting the spectra for the adopted host redshift of $z$=0.000804,
we take the average of the two lines to derive a Doppler velocity for
the Na~{\sc i} D absorption of +7 ($\pm$1) km s$^{-1}$, relative to
the adopted redshift of M101 (which is about $+$241 km s$^{-1}$).
When we interpret the observed velocities below in Section~\ref{sec:disc},
we take this value of $+$7 km s$^{-1}$ as representative of the
velocity of interstellar material in the vicinity of SN~2023ixf that
results from galactic rotation, and therefore as a likely indication of
the progenitor star's systemic velocity.  In
figures in this manuscript showing the observed line profiles, the velocity scale is only
corrected for $z$=0.000804, but we show this +7 km
s$^{-1}$ systemic velocity with a vertical green dashed line.

\begin{figure}
    \centering
     \includegraphics[width=3.1in]{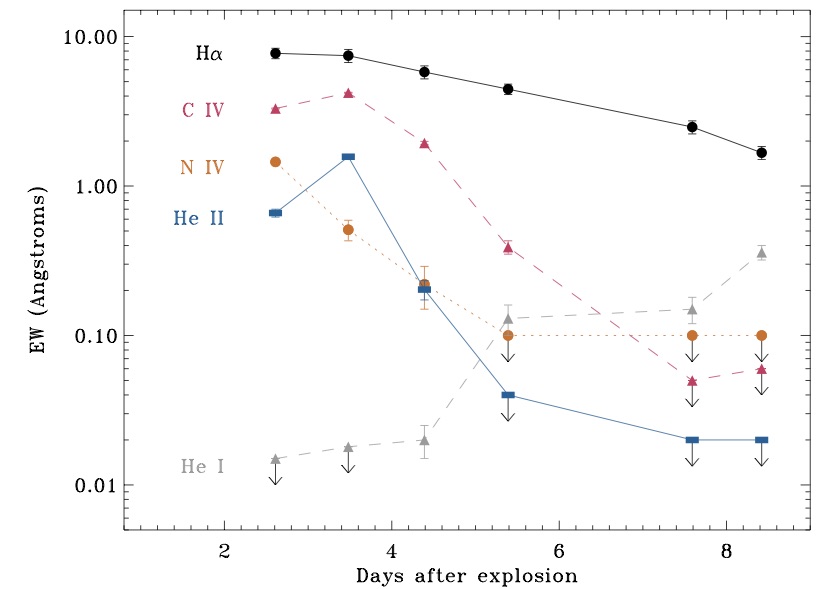}
   \caption{Measured equivalent widths (EW; emission lines have
     positive values here) for H$\alpha$, He~{\sc ii} $\lambda$4686,
     C~{\sc iv} $\lambda\lambda$5801,5811, He~{\sc i} $\lambda$5876,
     and N~{\sc iv} $\lambda\lambda$7109,7123.  The EW here is the integrated flux of the line (not a functional fit) divided by the continuum level, and the EW error bars are dominated by the noise in the nearby contiunuum level for these lines.}
   \label{fig:ew}
\end{figure}

\begin{figure}
    \centering
     \includegraphics[width=3.1in]{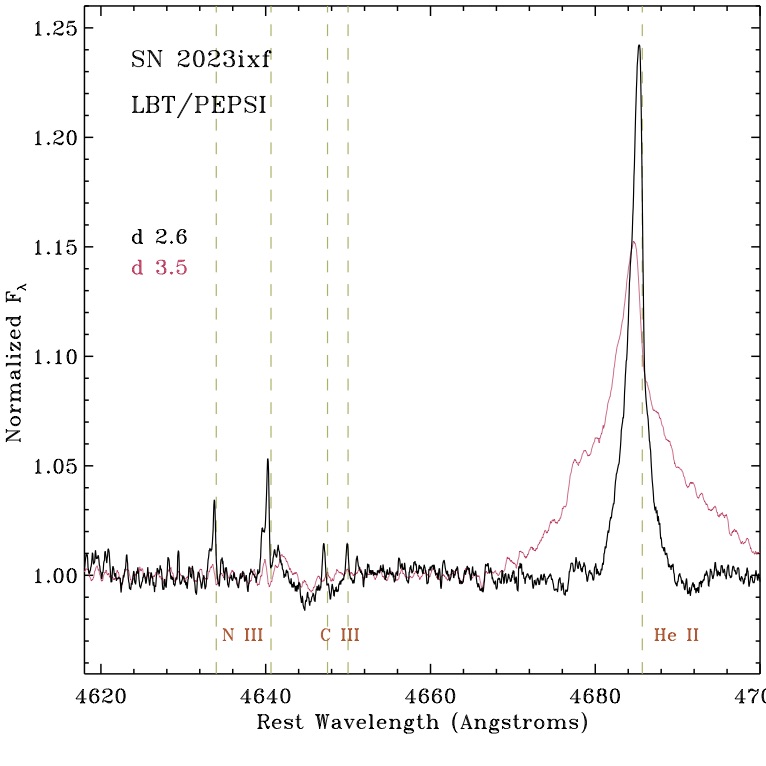}
   \caption{A detail of the spectrum in the wavelength range of the blue WR bump, including He~{\sc ii} $\lambda$4686, as well as the N~{\sc iii} $\lambda\lambda$4634,4641 and C~{\sc iii} $\lambda\lambda$4648,4650 lines.  For each line, the +7 km s$^{-1}$ systemic velocity is marked by a vertical dashed green line.  The N~{\sc iii} and C~{\sc iii} lines are only detected on our first epoch on day 2.6 where they appear weak, narrow, and blueshifted.}
   \label{fig:n3c3}
\end{figure}

\subsection{High ionization Features}\label{sec:ion}

Notable high ionization lines in these early-time echelle spectra are
He~{\sc ii} $\lambda$4686, C~{\sc iv} $\lambda\lambda$5801,5811, and
N~{\sc iv} $\lambda\lambda$7109,7123.  The line profile evolution of
each of these can be seen in Figures~\ref{fig:quad}a, \ref{fig:quad}b, 
and \ref{fig:quad}d, respectively (Panel \ref{fig:quad}c shows He~{\sc i} $
\lambda$5876, discussed in the next section).  If we take into
account the fact that C~{\sc iv} and N~{\sc iv} are closely spaced
doublets or blends, the evolution of strength and line profile shape
in each of these is similar.  All three have relatively strong, narrow
($\la$100 km s$^{-1}$), blueshifted ($-$50 to $-$150 km s$^{-1}$) emission peaks with broader wings
at the first epoch.  These lines then appear to broaden and fade over
the next 2-3 days, and completely disappear 3-4 days later.  The
evolution of the equivalent widths (EWs) of these emission lines
measured in our PEPSI spectra are shown in Figure~\ref{fig:ew}.  All
the high-ionization lines fade by factors of 10-20 or more over a time
period of 4 days.

The change in width of He~{\sc ii} over only 1 day
from May 21 to 22 is particularly stunning, where the broad $\pm$1000
km s$^{-1}$ emission wings come and go very quickly. The He~{\sc ii} profile on day 2.6 is narrow, lacking the broad electron scattering wings seen in H$\alpha$ on that same date (as we show below, when we subtract off a 1000 km s$^{-1}$ Lorentzian profile from H$\alpha$, the two line profiles are very similar).\footnote{We note that some lower-resolution spectra of SN~2023ixf do show broader wings of He~{\sc ii} $\lambda$4686 at very early epochs comparable to our observations \citep{wynn23}. Because PEPSI spectra are cross-dispersed, the ``continuum'' level that is normalized in the pipeline reduction may include some of the broadest emission wings if it stretches across more than one order.  We find this to be likely, so we caution the reader about the absence of broad He~{\sc ii} emission in these spectra, and we do not emphasize this result in our discussion.}    He~{\sc ii} then fades and disappears by day 5.4.    Of these
high-ionization lines, C~{\sc iv} $\lambda\lambda$5801,5811 has the
broadest wings (extending to $-$2000 km s$^{-1}$ on the blue side) and
it lingers the longest before fading, disappearing from the spectrum
about a day later than He~{\sc ii}. In any case, all narrow and
intermediate-width emission from these high-ionization features is
completely absent from the spectrum by day 7.6.  The narrow emission component fades more quickly, leaving
only a fainter intermediate-width component to persist for a few days.
High-ionization emission might fade either because the gas cools and
recombines, or because the CSM is overtaken by the SN; we return to
this topic later in Section \ref{sec:disc}.

There is a persistent weak emission feature on the blue wing of
He~{\sc ii} at about $-450$ km s$^{-1}$, possibly with a P Cygni
profile.  This is most likely an artifact that arises where edges of
echelle orders are merged (see the top panel of Figure~\ref{fig:full},
as noted earlier).  This is probably not emission from N~{\sc iii}
$\lambda\lambda$4679.4,4679.8, which is, however, seen in low-resolution spectra a day earlier (Bostroem et al., in prep.).

Notable for their general absence in our PEPSI spectra are N~{\sc iii} $\lambda\lambda$4634,4641 and C~{\sc iii} $\lambda\lambda$4648,4650.  Together with He~{\sc ii} $\lambda$4686, these lines constitute the so-called blue WR bump.  In many examples of SNe~II with fleeting CSM interaction signatures, these N~{\sc iii}/C~{\sc iii} lines are very strong (often equal in strength to He~{\sc ii} $\lambda$4686), and with strong electron scattering wings \citep{niemala85,leonard00,galyam14,smith11iqb,terreran22,wynn22}.  In fact, these N~{\sc iii}/C~{\sc iii} lines are seen in our spectra of SN~2023ixf, but only in our first spectrum on day 2.6, where they are extremely weak (Fig~\ref{fig:n3c3}).  
For N~{\sc iii} $\lambda\lambda$4634,4641 we measure EWs of 0.022 $\pm$0.004 and 0.035 $\pm$0.005 \AA, respectively, and for C~{\sc iii} $\lambda\lambda$4648,4650 we measure EWs of 0.013 $\pm$0.003 and 0.008 $\pm$0.002 \AA, respectively.  
The lines disappear the next day. Also, when seen on day 2.6, they only show the narrow emission components with no broad wings; these narrow components have the same blueshift and approximately the same width as He~{\sc ii} (Fig~\ref{fig:n3c3}).  These lines are stronger the previous day in lower-resolution spectra (Bostroem et al., in prep.), as noted above for N~{\sc iii} $\lambda\lambda$4679.4,4679.8.  Over the same period from day 2.6 to 3.5 when these lines vanish from our spectra, the strengths of C~{\sc iv} and He~{\sc ii} are still increasing (Fig.~\ref{fig:ew}).  This indicates that even as late as 2-3 days after explosion, the compact CSM is still increasing in ionization level, even though the light travel time to 20-30 AU is only about 3 hrs.  This suggests that a sudden flash of ionization from shock breakout is probably not the primary ionization source for the CSM, which may instead be photoionized by emission from the ongoing shock/CSM interaction \citep{smith11iqb,terreran22}.

\begin{figure}
    \centering
    \includegraphics[width=3.1in]{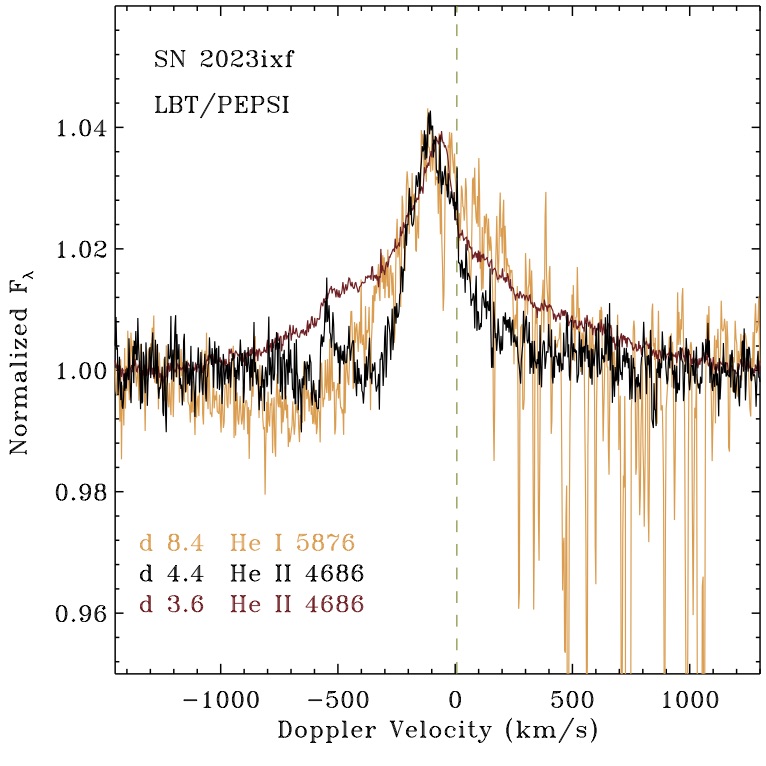}
   \caption{The line profile shape of the fading He~{\sc ii}
     $\lambda$4686 emission on days 3.6 (red) and 4.4 (black), as
     compared to the He~{\sc i} $\lambda$5876 emission a few days
     later (orange).  As He~{\sc ii} $\lambda$4686 emission
     wings within $\pm$1000 km s$^{-1}$ fade, He~{\sc i} $\lambda$5876
     emission and P Cygni absorption over the same range of
     wavelengths becomes stronger.  This suggests that the gas
     expanding at around 1000 km s$^{-1}$ (i.e. gas in the post-shock
     shell of swept-up CSM) is cooling and recombining.}
   \label{fig:helium}
\end{figure}

\subsection{Recombination of He II to He I}

The detailed evolution of He~{\sc i} $\lambda$5876 is shown in
Figure~\ref{fig:quad}c.  This line is often seen as a strong narrow
emission line in early SNe II with CSM signatures, and in SNe IIn, but
it is totally absent in the first two epochs of our echelle spectra of
SN~2023ixf.  Interestingly, however, He~{\sc i} $\lambda$5876 starts
to grow in strength and becomes an admittedly still very weak
intermediate-width emission feature by days 7.6 and 8.4 (blue and violet
in Figure~\ref{fig:quad}c; note that this line is found amid a forest
of telluric H$_2$O absorption features).  Figure~\ref{fig:ew} shows
the equivalent widths of He~{\sc i} emission as compared to other
lines, demonstrating how the strength of He~{\sc i} emission increases
as He~{\sc ii} and the other high-ionization lines fade away over 4-5
days. (Note that while the actual value of the EW for He~{\sc i} is
quite uncertain because of all the overlapping telluric absorption, the
relative increase in strength of He~{\sc i} shown in
Figure~\ref{fig:ew} is real because the forest of telluric lines does not alter the flux passing between them.)
The line profiles of He~{\sc i} during this evolution can be seen in
Figure~\ref{fig:naid} as noted earlier, but are shown more clearly in
Figure~\ref{fig:quad}c.  The emission component of He~{\sc i} has a width of about
500-1000 km s$^{-1}$; there is no narrow emission from He~{\sc i}
$\lambda$5876 that would correspond to the blueshifted, narrow (50-100
km s$^{-1}$) peak of He~{\sc ii} in the first epoch.  He~{\sc i}
$\lambda$5876 shows weaker intermediate-width P Cyg absorption, discussed more below.

This lack of narrow emission indicates that the weak He~{\sc i}
$\lambda$5876 emission detected in these spectra is arising from
accelerated gas that has already been swept up by the shock and is now
cooling and recombining.  Accordingly, this indicates that the narrow
component of He~{\sc ii} (and by extension, the narrow components of
N~{\sc iv} and C~{\sc iv}) disappear because the slow gas is
accelerated by the shock, not because it survives as pre-shock CSM
and recombines.

\begin{figure*}
    \centering
    \includegraphics[width=6.0in]{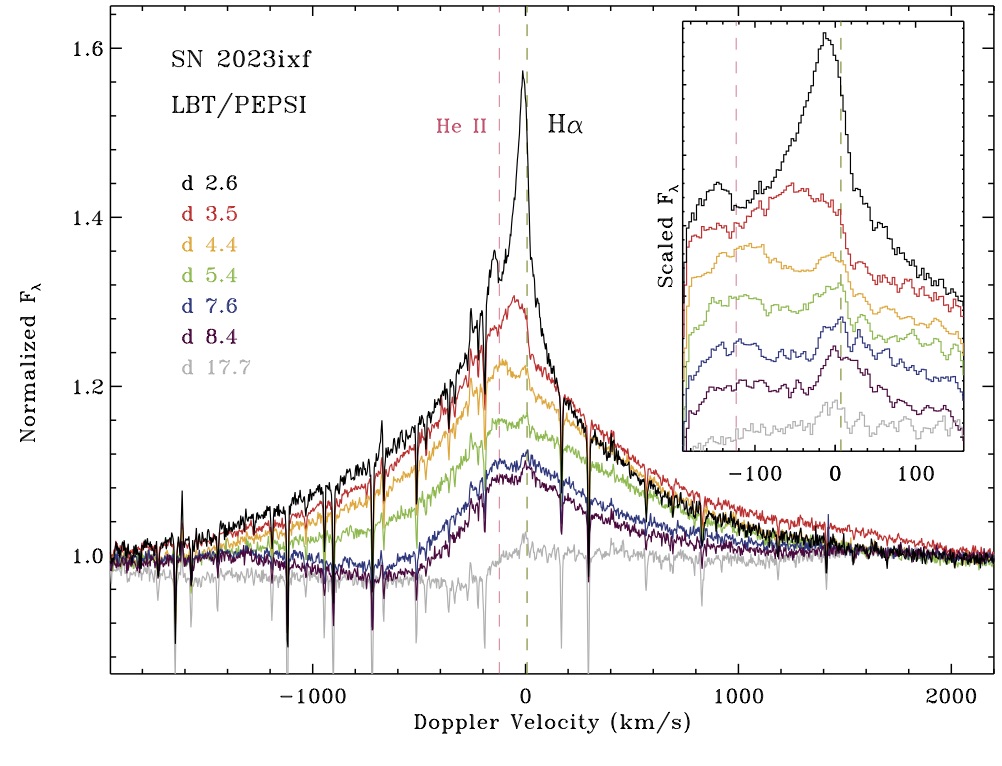}
   \caption{Same as Figure~\ref{fig:quad}, but for the H$\alpha$ line
     profile.  The inset in the upper right zooms-in on the narrow
     component.  Note that the emission bump at around $-$130 km
     s$^{-1}$ is not a velocity component of H$\alpha$, but is instead
     emission from the weaker He~{\sc ii} $\lambda$6560 line that
     overlaps (its +7 km s$^{-1}$ systemic velocity is marked with a magenta vertical
     dashed line).  In this case, it would therefore be incorrect to interpret the
     drop in flux between this feature and the narrow H$\alpha$
     emission peak as narrow P Cygni absorption of H$\alpha$.}
   \label{fig:ha}
\end{figure*}

Figure~\ref{fig:helium} compares profiles of the fading He~{\sc ii}
$\lambda$4686 emission to the last epoch of He~{\sc i} $\lambda$5876
emission.  Note that as He~{\sc ii} $\lambda$4686 emission wings
within $\pm$1000 km s$^{-1}$ fade away, the He~{\sc i} $\lambda$5876
emission and P Cygni absorption over the same range of velocities
become stronger.  This suggests that the gas expanding at around 1000
km s$^{-1}$ is cooling and recombining.  These velocities are
significantly faster than the narrow (50-100 km s$^{-1}$) component
from the unshocked CSM seen in the first-epoch spectra.  Again, this
confirms that the He~{\sc i}-emitting gas has been accelerated,
probably because it is now in the post-shock shell of swept-up CSM.
The He~{\sc i} $\lambda$5876 profile on day 8.4 also shows a
weak and broad P Cygni absorption feature at $-$500 to $-$1200 km
s$^{-1}$, which is similar to the later epochs of H$\alpha$ discussed
below.

Interestingly, we do not see a similar ionization transition in alpha
elements.  While C~{\sc iv} and N~{\sc iv} fade quickly over a few
days, we do not see a corresponding growth in the strength of N~{\sc
  iii} or C~{\sc iii} lines that are seen in several other SNe II with
early narrow CSM features, as noted above.  At the end of our spectral series, the C~{\sc iii} and N~{\sc iii}
emission features do not turn on as the C~{\sc iv} and N~{\sc iv}
fade.  This implies that the C~{\sc iv} and N~{\sc iv} is not fading
primarily because the N and C ions are recombining to a lower
ionization state.  Instead, it may suggest that the CSM and shocked
shell are largely getting enveloped by the expanding SN photosphere
after a week; this is discussed more below.

\begin{figure}
    \centering
    \includegraphics[width=3.2in]{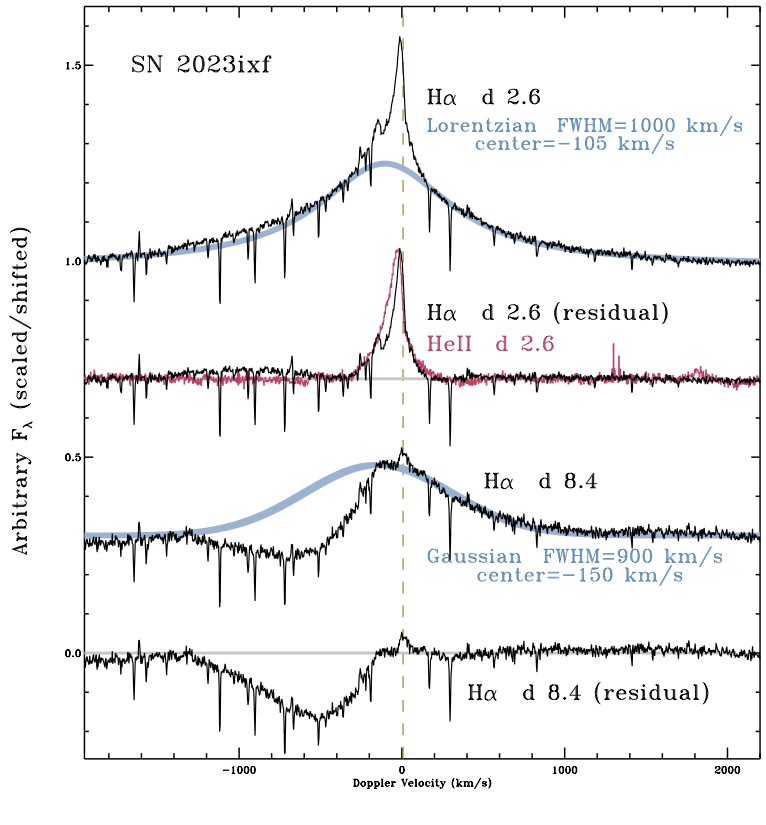}
   \caption{Details of the H$\alpha$ profile.  The top spectrum is the
     H$\alpha$ profile observed on our first epoch on day 2.6,
     compared to a Lorentzian function with FWHM=1000 km s$^{-1}$ and
     a centroid shifted to the blue by 105 km s$^{-1}$.  This gives an
     adequate match to the red wing, but not the central peak or blue
     wing.  The residual after subtracting this Lorentzian curve is
     shown in the second tracing from the top (black), which is then
     compared to the observed He~{\sc ii} $\lambda$4686 profile on the
     same date (plotted in magenta).  The third spectrum down shows
     the observed H$\alpha$ profile on day 8.4
     compared to a Gaussian with FWHM=900 km s$^{-1}$ that has
     its centroid shifted 150 km s$^{-1}$ to the blue.  Finally, the
     bottom tracing shows the residual after subtracting this
     Gaussian, highlighting the shape of the blueshifted absorption
     feature.}
   \label{fig:hafits}
\end{figure}

\subsection{Intermediate-width and Narrow H$\alpha$}

H$\alpha$ exhibits the most interesting and informative evolution of
the lines seen in the spectrum during the first week after explosion
(Fig.~\ref{fig:ha}).  Overall, H$\alpha$ displays a clear and steady
evolution from a narrow line core with broad Lorentzian-shaped wings
at the first epoch, transitioning to a clear intermediate-width P
Cygni profile a week later.  It is similar to the evolution seen in
normal SNe IIn, but on a vastly compressed timescale. The narrow
component fades more rapidly than the intermediate-width component.
Overall, the EW of H$\alpha$ fades by about a factor of 5 in the first
week (Fig.~\ref{fig:ew}), during a time when the $r$ magnitude only
brightens modestly \citep{griffin23ixf}, corresponding to a factor
of $\sim$1.6 increase in continuum flux.  This indicates that the
narrow/intermediate H$\alpha$ line luminosity fades by a factor of
$\sim$3 during the first week of observations.  By 2 weeks after explosion (day 17.7 in Fig.~\ref{fig:ha}), the intermediate-width emission component of H$\alpha$ is gone.  Some of the details of
the line profile evolution are interesting.  Note that H$\alpha$ is blended with weak emission from He~{\sc ii} $\lambda$6560, which produces a small bump of excess emission on the blue wing of H$\alpha$ at $-$130 km s$^{-1}$; this is discussed at the end of the current section.

First, we consider the evolution of the intermediate-width component.
Aside from an overall fading with time, the shape of the red wing of
the line changes little, with a gradual reduction in maximum velocity
from about $+$2,000 km s$^{-1}$ on the first epoch down to about
$+$1,000 km s$^{-1}$ a week later.  Some of this apparent slowing may
simply arise because the extremes of the line wings fade below the
noise, but it may also result from cooling of the region where the
electron scattering occurs.

There are more dramatic and important changes occurring on the blue
wing of the line.  At first the blue wing appears as a nearly
symmetric reflection of the red wing, extending to $-$2,000 km
s$^{-1}$.  However, the blue side fades more quickly than the red
wing, steadily transforming into an intermediate-width P Cygni
absorption feature with a trough at $-$700 km s$^{-1}$ and a blue edge
at about $-$1,300 km s$^{-1}$.  This change is physically significant.
The intermediate-width wings of interacting SNe are often presumed to
be caused by thermal electron scattering of the narrow line emission,
broadening those narrow line photons into wings that can be
approximated by a Lorentzian shape
\citep{chugai01,smith08,smith17review}.  This is the case for our
first spectrum on day 2.6 (Fig.~\ref{fig:hafits}).  A Lorentzian with a
FWHM=1000 km s$^{-1}$ and with a center shifted 105 km s$^{-1}$ to the
blue matches the line wing shapes of H$\alpha$ on day 2.6 reasonably
well, except for some low-level broad excess emission on the blue
wing.  Electron scattering is thermal, and the wings are expected to
be symmetric about the wavelength of the original narrow-line photons.\footnote{Although note that the original velocity of the line photons that were scattered and therefore define the center of the Lorentzian (at $-$105 km s$^{-1}$) is not necessarily the same as the velocity of the observed narrow line photons that have escaped without scattering ($-$22 km s$^{-1}$) if there is asymmetry, asymmetric illumination of the CSM, or a velocity gradient in the CSM; this is likely to be the case, as we discuss later in Section 4.  However, the observed narrow emission does span a range of velocity that includes the centroid of the Lorentzian.}  Therefore, the $-$105 km s$^{-1}$ blueshifted centroid of this 1000 km s$^{-1}$ Lorentzian makes sense in this case, since the narrow emission that is being scattered and broadened spans a similar range of velocity.

However, electron scattering cannot turn narrow emission into a broad
absorption feature.  The transition of SN~2023ixf's H$\alpha$ line to
an intermediate-width P Cygni absorption feature requires that after day 3, we are
seeing Doppler-shifted absorption and emission by accelerated H atoms in the post-shock gas,
and not emission from slow moving pre-shock gas that has been broadened
mostly by electron scattering (although electron scattering may
obviously still influence the red wing of the line, for example).

The fact that this broader absorption increases in strength over a
time period when the narrow emission component mostly fades away
strongly suggests that after a few days, most of the pre-shock CSM has
been swept up by the shock.  After that, H$\alpha$ emission and
absorption traces CSM that has been hit by the forward shock and
accelerated, being swept up into a dense post-shock shell (often
called the ``cold dense shell'', or CDS, in SNe IIn).  By day 8.4, the
red wing of the line can no longer be well matched by a Lorentzian
shape; instead, a Gaussian function with FWHM=900 km s$^{-1}$ and with
its center shifted to the blue by 150 km s$^{-1}$ gives a better match
(Fig.~\ref{fig:hafits}).  The fact that the CDS is seen in absorption
against the SN continuum photosphere requires that it has indeed
cooled.  Similar intermediate-width absorption features arising in the
CDS are seen in some SNe IIn, including SN~2006gy \citep{smith10gy}
and SN~1994W \citep{chugai04}.  We also note that the width and shape
of the H$\alpha$ P Cygni profile on day 8.4 is very similar to He~{\sc
  i} $\lambda$5876 on the same date (Fig.~\ref{fig:helium}).

Although the intermediate-width H$\alpha$ persists longer than the high-ionization lines, it doesn't last long.  Figure~\ref{fig:ha} also shows the observed H$\alpha$ profile in a PEPSI spectrum on day 17.6 (gray).  While there is a gap in our spectral coverage, this shows that the intermediate-width emission component of H$\alpha$ is gone by a little over 2 weeks after explosion. There is still a kink at around zero velocity, hinting at some persistent intermediate-width P Cygni absorption of H$\alpha$ at this time.  However, there is also a deficit of flux at high velocity (resembling a lower continuum level on the blueshifed side of the line on day 17.6; Fig~\ref{fig:ha}), suggesting that the broad absorption from SN ejecta is beginning to influence the spectra by this epoch.

The narrow component of H$\alpha$, presumably arising from the
pre-shock CSM, shows surprisingly complex profile evolution.
Excluding the effects of broadening from electron scattering discussed
above, one might expect to see a very narrow (10-20 km s$^{-1}$) and
symmetric profile shape from the core of an emission line arising from
a more-or-less spherical and slow RSG wind, but this is evidently not
the case in SN~2023ixf.

The inset (upper right) in Figure~\ref{fig:ha} documents changes in
line profile shape of the narrow component of H$\alpha$ (see also Fig.~\ref{fig:hafits}).  At our first
epoch (day 2.6; black), the narrow component is asymmetric (a broader
blue wing and sharper drop on the red side), it has a blueshifted
centroid (at about $-$25 km s$^{-1}$), and it has a FWHM of 48 km
s$^{-1}$.  At the second epoch on day 3.5 (red) it becomes weaker,
broader (FWHM = 79 km s$^{-1}$), and even more blueshifted (centroid
velocity of $-$42 km s$^{-1}$).  In both of these first two epochs,
the narrow component seems rather abruptly cut off at zero velocity
on the red wing.  After that, the narrow component becomes much
weaker, and settles down to a more symmetric and narrower (FWHM
$\approx$ 45 km s$^{-1}$) emission component that has a centroid
closer to zero velocity or even slightly redshifted (about +8 km
s$^{-1}$).  At all epochs, the FWHM of the narrow component is
resolved, being significantly larger than the instrumental resolution
of about 6 km s$^{-1}$.

How shall we interpret the changing offsets in the centroid velocities
of the narrow emission component?  In Section 3.1, we noted that the
centroid velocities of the interstellar Na~{\sc i} D absorption (after
correction for M101's redshift of $z$=0.000804) was $+$7 km s$^{-1}$,
which we take to be the likely systemic velocity of the progenitor.
This agrees (to within 1 km s$^{-1}$) with the centroid velocity that
we measure for the narrow H$\alpha$ component at $\sim$1  week post-explosion.  One possible interpretation of this is
that the lingering narrow emission centered on the systemic velocity
with a resolved width of 45 km s$^{-1}$ corresponds to photoionized
gas in distant regions of the progenitor's RSG wind, which may be
roughly spherical, or at least symmetric about the systemic velocity.
This, in turn, means that the pronounced blueshift of the narrow
H$\alpha$ component on days 2-3 (as well as the similar blueshift
of narrow components of He~{\sc ii} and other high ionization lines on
these same dates) is real.  Thus, in early epochs, we are seeing
emission from inner regions of the CSM that are predominantly on the
near side of the SN, which are expanding toward us at 30-50 km
s$^{-1}$ or more (the observed blueshifts are even larger for higher
ionization lines).  We return to the implications of this blueshifted
narrow emission later in Section~\ref{sec:disc}.

At no time during the first week do the spectra show any hint of a
narrow P Cygni absorption component in H$\alpha$ that might arise from
absorption along the line of sight through dense, slow, pre-shock CSM.
Such narrow absorption features from the pre-shock CSM are often seen
in SNe IIn, providing that spectra have sufficient resolution
\citep[e.g.][]{salamanca02,trundle08,chugai19,smith10gy,smith20}. We
note that the narrow emission bump seen at $-$130 km s$^{-1}$ is
actually weak narrow emission from He~{\sc ii} $\lambda$6560
superposed on the blue side of the H$\alpha$ line.  It is not a
separate velocity component of H$\alpha$, and the gap between these
two is not narrow P Cygni absorption of H$\alpha$.  The expected systemic
velocity of this He~{\sc ii} line is marked with a vertical
dashed magenta line in Figure~\ref{fig:ha}.  The emission feature in question is blueshifted
from this reference position by about $-$30 km s$^{-1}$, similar to
the blueshift of the narrow components in other lines at the same
early epoch.  The lack of narrow P Cyg absorption may suggest that the CSM is asymmetric, as discussed below.

\begin{figure}
    \centering
    \includegraphics[width=3.2in]{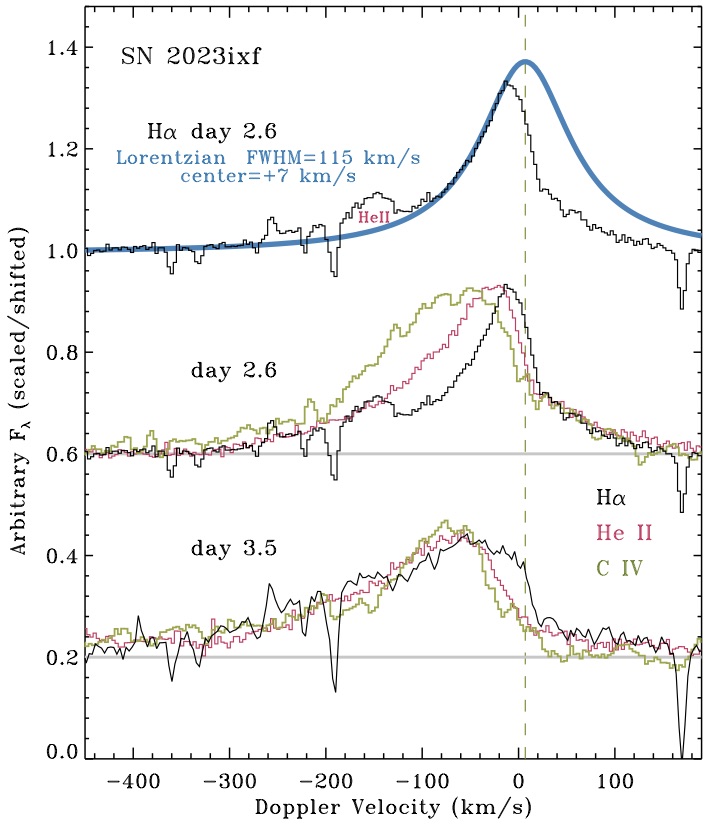}
   \caption{Details of the narrow-line profiles in our first two
     epochs of PEPSI spectra.  The top spectum here is the same as the second spectrum from the top shown in
     Figure~\ref{fig:hafits}.  This is the narrow H$\alpha$ profile on
     day 2.6, which is the residual flux remaining after subtracting
     away a broad Lorentzian with FWHM=1000 km s$^{-1}$.  For
     comparison, a symmetric narrow Lorentzian function with FWHM =
     115 km s$^{-1}$ is shown in blue, centered at the systemic
     velocity of $+$7 km s$^{-1}$ (indicated by the green
     vertical dashed line).  This Lorenztian function matches the blue
     wing of the line, and is meant to illustrate what the full
     intrinsic symmetric narrow H$\alpha$ emission profile from the CSM would
     look like if the red side of the line were not missing.  Recall
     that the bump at $-$130 km s$^{-1}$ is He~{\sc ii} $\lambda$6560.
     The bottom two plots show tracings of the narrow components of
     H$\alpha$ (black), He~{\sc ii} $\lambda$4686 (magenta), and
     C~{\sc iv} $\lambda$5801 (green) overplotted, on day 2.6 (middle)
     and day 3.5 (bottom).  As with the top plot, each of these has
     had a broad Lorentzian subtracted in order to highlight the
     narrow emission.}
   \label{fig:narrow}
\end{figure}

\section{Discussion} \label{sec:disc}

\subsection{Blueshifted Narrow features and the CSM Speed}

As noted above, all of the emission lines seen in our first
PEPSI epoch on day 2.6 have narrow emission peaks that are blueshifted
(see Figures~\ref{fig:quad} and \ref{fig:ha}).  Recall that the
presumed systemic velocity of SN~2023ixf (indicated by
Na~{\sc i}~D absorption) is at $+$7 km s$^{-1}$ relative to the average
redshift of M101.  Relative to the systemic velocity, the narrow
H$\alpha$ peak is at $-$22 km s$^{-1}$ and the narrow emission
component extends from about $+$15 km s$^{-1}$ at the drop on the red
side, reaching out to about $-$100 km s$^{-1}$ or more on the blue
side.  Similarly, the high ionization lines (He~{\sc ii}, C~{\sc iv},
N~{\sc iv}) have their strongest narrow emission on day 2.6 ranging
from about 0 km s$^{-1}$ to $-$150 or $-$200 km s$^{-1}$.

Thus, it appears that the narrow emission on day 2.6 is being emitted by
expanding CSM that is primarily on the near (approaching) side of the
SN.  This has been seen before in early spectra of interacting SNe,
and seems to be well understood as being due to the combined effects
of light travel time in the presence of a variable illumination source
and occultation by the SN photosphere \citep{groh14,shivvers15,gv16}.
This does not mean that the CSM is actually one sided, but that we are
not able to detect as much emission from the redshifted side of the
CSM because it is blocked from our view or not yet fully illuminated,
depending on the time after explosion.  If much of the redshifted CSM
emission is missing, this is important for interpreting the
velocities.

Figure~\ref{fig:narrow} shows a detail of the narrow H$\alpha$
component in our first spectrum, after subtracting away a broad
Lorentzian as in Figure~\ref{fig:hafits}.  This is compared to a
narrow Lorentzian function centered at the systemic velocity of $+$7
km s$^{-1}$, matching the blue wing of H$\alpha$ (note that this curve
does not match the excess emission bump at $-$130 km s$^{-1}$, which is
due to emission from He~{\sc ii} $\lambda$6560 superposed on the
H$\alpha$ line; with this assumed line center at +7 km s$^{-1}$, the uncertainty in the Lorentzian FWHM is about $\pm$10 km s$^{-1}$).  This symmetric Lorentzian curve demonstrates what
the true symmetric narrow H$\alpha$ emission from the CSM would look
like at this epoch, if we could detect all of it.  This, in turn, tells us
that the CSM expansion speed indicated by the narrow H$\alpha$
profile is actually about 115 km s$^{-1}$.  This is much faster than a typical RSG
wind of $\sim$20 km s$^{-1}$.

However, interpreting this blueshifted emission is complicated by the
fact that the amount of blueshift is different for different lines (tracing different ionization levels),
and the amount of blueshift in the narrow components is time
dependent.  This is discussed next.

\subsection{Acceleration of the pre-shock CSM}

Figure~\ref{fig:narrow} illustrates the complex time-dependent and
ionization-dependent evolution of the narrow emission components from
the pre-shock CSM in our first two epochs of PEPSI spectra (these are
day 2.6 and 3.5, shown in the middle and bottom tracings in
Fig.~\ref{fig:narrow}, respectively).  Note that we have removed the
broader wings of these profiles by subtracting Lorentzian functions
with FWHM values of around 1000 km s$^{-1}$ (as shown for H$\alpha$ in
Fig.\ref{fig:hafits}). Figure~\ref{fig:narrow} includes narrow profiles
of H$\alpha$ (black), He~{\sc ii} (magenta), and C~{\sc iv} (green),
which trace recombination emission from increasing ionization levels
(13.6~eV, 54.4~eV, and 64.5~eV, respectively).
Figure~\ref{fig:narrow} does not include N~{\sc iv} because of its
relatively low signal to noise.

On day 2.6 (middle tracings), all the narrow lines are blueshifted,
but there is a remarkable and systematic increase in both width and in
the amount of blueshift of the line center as we move from H$\alpha$
to He~{\sc ii} and then to C~{\sc iv}.  Basically, higher ionization
lines exhibit faster outflow speed.  There are two potential
explanations for this.


1. The source function for each of these lines may have a slightly
different radial dependence in an ionized pre-shock CSM, because of
radial gradients in CSM density and ionization level
\citep[e.g.,][]{groh14}.  For example, one might expect higher ionization lines to arise closer the shock, and H$\alpha$ to be emitted from a more distant region. If there is a velocity gradient in the CSM
(with faster velocities at smaller radii), this might account for why
higher ionization lines show faster outflow speeds.  However, this
would require a very steep velocity gradient (increasing by a factor
of more than 2 from H$\alpha$ to C~{\sc iv}) in a narrow radial zone.
In the models presented by \citet{groh14}, the CSM velocity was
assumed to be constant with radius, but new models may be able to
quantify the velocity gradient that would be needed to explain the
observed profiles.  The required velocity gradient is not only steep,
but decreases with radius (faster outflow at smaller radii).  This is
the opposite of what is expected for a Hubble-like flow from a sudden burst of
pre-SN mass loss, and it is the opposite of an acceleration zone of a stellar wind that eventually reaches its terminal speed at a large radius.  Instead, it would point to either dramatic radiative
acceleration of the inner pre-shock CSM by the radiation from the shock
itself, or to an outflow speed that was increasing during the few years preceding explosion.  If the CSM is very dense, a small mean free path could lead
to a sharp velocity gradient immediately ahead of the shock, and this
would be an interesting problem for detailed models to quantify.  On the other hand, a rapidly changing outflow speed in the pre-SN ejection is perhaps plausible but is {\it ad hoc}.

2.  Another possibility is that the CSM is asymmetric.  Imagine that
the SN blast wave hits CSM with a disk or torus geometry that has a
dropping density away from the equatorial plane.  In the equatorial
plane, the forward shock will hit the densest CSM and will be
decelerated the most.  Moving out of the plane, the shock will
encounter relatively less dense CSM, and hence, the forward shock will
decelerate less and will continue outward at a higher speed.  At these
intermediate latitudes, the faster shock will yield hotter post-shock
gas, and so the CSM immediately ahead of the shock will be illuminated
by a harder radiation field from the shock.  Thus, we might expect to
see higher ionization tracers (i.e. C~{\sc iv}) coming preferentially
from higher latitudes in the immediate pre-shock environment, and
lower ionization (H$\alpha$) emitted by CSM in the dense equatorial
plane.  The different velocities at different ionization might then
arise because the less dense CSM out of the plane also has less
inertia, and will therefore experience more radiative acceleration, or because gas at mid-latitudes was ejected with higher speeds than gas in the equator, as commonly seen in bipolar nebulae.
This is a difficult scenario to investigate theoretically, because it
requires 3-D radiation hydrodynamics with line transfer.  However, if
the pre-shock mean free path is small at these high densities, one
might address the question with a series of spherical calculations
with shocks running into a range of different CSM density.

So far we have only discussed the narrow profiles in our first epoch
on day 2.6; the bottom tracings in Figure~\ref{fig:narrow}
show that the situation changes markedly one day later on day
3.5.  Namely, the stark differences in velocity in the three
different lines are mostly gone.  While the red edge of the narrow
component of H$\alpha$ still extends closer to zero, the width and
blueshifted centroid velocity are now much more similar for the three
lines.  The essential change is that most of the narrow peak of H$\alpha$ is gone, making the H$\alpha$ line appear broader and
more blueshifted (see the inset of Figure~\ref{fig:ha}), and He~{\sc ii} as well to a lesser degree, making
all three lines similar.  The factor of 2 difference in velocity
between H$\alpha$ and C~{\sc iv} is now gone.  As above, there are again two
potential explanations for this change:

1.  We may be seeing direct evidence for the pre-shock radiative
acceleration of CSM.  After more time has passed, the radiation field
has accelerated the rest of the slower CSM traced by H$\alpha$
emission so that it now shares the faster expansion speeds seen in
higher ionization gas.  Detailed models would be needed to confirm that the timescale and amount of acceleration (as well as the difference between H$\alpha$ and the higher ionization lines) is plausible.

2.  Another possibility arises if the CSM is asymmetric. As noted
earlier, the slowest and densest CSM in the equatorial plane that
emitted the narrow peak of H$\alpha$ would be at a smaller radius
because the blast wave in this direction gets decelerated the most.
As such, the dense and slow equatorial CSM at the narrow pinched waist
would be the first to be enveloped by the expanding SN photosphere.
In this scenario, the H$\alpha$ line would appear broader simply
because the narrowest emission is engulfed and hidden by the SN ejecta and
therefore removed from the observed line profile, not because that slow
gas is accelerated.  This would be consistent with the fact that the
flux of the narrow component of H$\alpha$ drops precipitously in this
first day, and with the apparent change in shape of the narrow H$\alpha$ line (again, see the d 2.6 and d 3.5 profiles in the inset of Figure~\ref{fig:ha}).

One difference that may help discriminate between these two
possibilities concerns the relative density of the emitting CSM.  In
the spherical case (option 1 for both epochs discussed above), the
narrower H$\alpha$ emission comes from lower ionization zones at
larger radii, and therefore should have lower densities compared to
the He~{\sc ii} and C~{\sc iv} emitting zones.  In the aspherical
second case, the narrowest H$\alpha$ is emitted by the densest CSM at
the equator, while the higher speeds at higher ionization result from
lower densities at higher latitudes.  Perhaps future modeling of the
spectrum can help quantify the physical conditions in the emitting
zones. Although it is tempting to ascribe the increasing H$\alpha$
speed from day 2.6 to day 3.5 to radiative acceleration of the CSM, it is
difficult to confirm this with available information because asymmetry
(which might be expected anyway) also provides a suitable and perhaps more straightforward explanation.

\subsection{Disappearing narrow lines and CSM radius}

Regardless of the details discussed above, the fact that the narrow
emission lines disappear after a few days --- a defining
characteristic of this class of events --- provides a straightforward
way to estimate the radial extent of the dense CSM.  Combined with
empirical expansion speeds, this can inform the timescale for the pre-SN mass
loss by the progenitor.  Whether the CSM is obliterated by the forward
shock or enveloped by the SN photosphere, the narrow lines should
disappear from the spectrum.

Narrow lines disappear after 1-2 days in our PEPSI
spectra, and the intermediate-width lines disappear after 3-4 days.
We assume that our first spectrum on May 21 was taken 2.6 days after
explosion, so this means that the narrow lines disappear in 3.6-4.6
days, and the intermediate-width components (except H$\alpha$)
disappear after 5.6-6.6 days.  These timescales can be used to
constrain the properties of the CSM.

In these early phases, we adopt an expansion speed for the SN ejecta
photosphere of 10,000 km s$^{-1}$.  The speed may be a bit slower as
time goes on, and the fastest ejecta are faster than this, but this
assumption is sufficient for a rough estimate.  At this speed the
relevant radius corresponding to the disappearance of the narrow
components on day 3.6 is $R$ = $vt$ = (3-5)$\times$10$^{14}$ cm or
about 20-30 AU.  This is only 5-10 $R_*$ for a typical RSG.  The
corresponding timescale of the pre-SN mass ejection depends, of
course, on the expansion speed of the CSM; i.e. $t_{\rm preSN} =
t_{\rm obs} \times (v_{\rm CSM} / v_{\rm SN})$.  For CSM produced by
a normal RSG wind speed of 10-20 km s$^{-1}$, the CSM would have been
ejected 5-10 years before explosion.  For a faster CSM expansion speed
corresponding to the observationally inferred expansion speed of 115
km s$^{-1}$ (see Fig~\ref{fig:narrow}), the pre-SN timescale for mass
ejection is more like 0.9-1.5 yr.  

Which one of these is correct returns us
to the ambiguous question of whether the CSM is aspherical and if it has
already been radiatively accelerated by the time of our first PEPSI
spectrum on day 2.6.  A timescale of 5-10 yr, expected for a normal
slow RSG wind speed, does not match any expected timescale for late
nuclear burning, but it is similar to the observationally inferred
timescales for some SNe IIn \citep{smith17review}.  We should not,
however, necessarily expect a phase of extreme pre-SN mass loss to
behave like a normal RSG wind.  On the other hand, a pre-SN ejection
timescale of $\sim$1 yr, derived from the observed CSM speed, agrees
well with the expected time for Ne or O burning.  This agreement may
favor instabilities in late Ne or O burning as the culprit for
triggering the extreme pre-SN mass loss for SN~2023ixf.  However, recent studies suggest that wave driving on its own may be unlikely to drive severe mass loss, instead being more likely to inflate the star's envelope \citep{lf20}.  As proposed by \citet{sa14}, however, this type of sudden pre-SN swelling may trigger severe and asymmetric mass loss in binary systems.

\subsection{Intermediate width features: Fading away, not broadening}

As noted above, the intermediate-width components of emission last
longer than the narrow lines, which fade after 1-2 days.  This is
especially true of H$\alpha$, which persists for at least a week after 
explosion as a strong, intermediate-width P Cygni line.  The 
intermediate-width emission component of H$\alpha$ disappears after 
another week or so, being gone by day 17.6 (gray spectrum in 
Fig.~\ref{fig:ha}).  Unlike H$\alpha$, the 
intermediate-width components of the higher ionization lines 
(He~{\sc ii}, C~{\sc iv}, N~{\sc iv}) do not develop any P Cygni 
absorption features before they vanish after a few days.  Why does this happen?

One reason for high-ionization lines to fade is recombination of the
gas.  We noted above, however, that as the N~{\sc iv} and C~{\sc
  iv} lines fade and disappear, we do not see a corresponding increase
in strength of N~{\sc iii} and C~{\sc iii} lines.  Also, although He~{\sc i} $\lambda$5876 (only seen as an intermediate-width component) does increase in strength as He~{\sc ii} fades, it never gets very strong, and it has disappeared again by day 17.7 (Fig.~\ref{fig:quad}b). Thus, even though
we do see some evidence of recombination and cooling in the post-shock
CDS (from H$\alpha$ P Cygni absorption and the increasing He~{\sc i}
emission), recombination is not the primary explanation for the
disappearance of the N~{\sc iv} and C~{\sc iv} lines.

Another reason why the intermediate-width emission from the post-shock
shell might fade would be if the SN ejecta that feed the reverse shock
are able to accelerate the shock front, essentially obliterating the
slower moving post shock gas as it is swept up to become part of the
fast SN ejecta moving at 5,000-10,000 km s$^{-1}$. In this case, we
would expect the intermediate-width lines to broaden from 1,000 to
5,000 km s$^{-1}$ or more as they fade.  This is not what the
observational data show either.  Instead, the intermediate-width components
stay at about the same width or even become slightly narrower as they
fade away.  Importantly, on days 7.6 and 8.4 when the high-ionization
lines have vanished, the P Cygni absorption seen in H$\alpha$ and
He~{\sc i} still maintains the same slower speeds of 500-1,300 km
s$^{-1}$.  This directly contradicts the idea that the post-shock shell 
is getting faster.  We therefore find it unlikely that the CSM interaction
signatures fade because the post shock shell is accelerated and
incorporated into the SN ejecta.

One last possibility arises if the CSM is significantly asymmetric, as
in a case where the CSM is primarily equatorial.  As noted above, the
shock front that crashes into the densest material in the equator will
be decelerated by the CSM.  However, in other directions with much
less dense CSM, the SN ejecta will expand unimpeded.  Since the CSM
here is found at radii of 20-30 AU, whereas the SN photosphere will
eventually reach a radius of around 100 AU, the slower CSM
interaction regions in the equatorial zones can be engulfed by the SN
ejecta and hidden inside the SN photosphere.  The opaque SN ejecta will 
wrap around the disk, if the disk is slow and thin enough.  Even if the 
SN ejecta do not completely engulf the disk, it may be hidden from observers, as 
discussed in detail previously by \citet{smith11iqb}, and invoked as
the explanation for the bizarre behavior of iPTF14hls \citep{as18}.
In this scenario, the CSM interaction zone with its relatively slow
CDS may still be there, but its emission is blocked from our view or
completely thermalized by the surrounding optically thick SN ejecta.
There are observed cases where the CSM interaction persists while it
is hidden beneath the photosphere, and signatures of strong CSM
interaction reappear when the recombination photosphere recedes (i.e. after the plateau drops), as in
PTF11iqb, iPTF14hls, SN~1993J, and SN~1998S
\citep{smith11iqb,as18,pozzo04,matheson00,smith17}. On the other hand, if the
asymmetric CSM has a low-enough total mass, it may indeed get
obliterated and incorporated into the fast SN ejecta during the time
when it is hidden beneath the photosphere, yielding little or no
lingering CSM interaction signatures after the photosphere recedes.
It will be interesting to see what happens after SN~2023ixf fades from
its plateau and/or enters it nebular phase in a few months.  

Note that this scenario where the CSM interaction region is engulfed by the SN photosphere only works of the CSM is highly asymmetric.  Of the various pre-SN ejection 
mechanisms related to late-phase nuclear burning mentioned in the 
introduction (wave driving, super-Eddington  winds, etc.), only 
pre-SN binary interaction triggered by envelope inflation \citep{sa14} 
is necessarily expected to produce strong asymmetry in the CSM.  This may turn out to be an important clue to the pre-SN mass loss.

Overall, the observational data seem to favor a scenario where the
narrow lines fade and disappear mostly because the slow-moving CSM is
swept up by the shock and accelerated, or even occulted.  The intermediate-width
components that are emitted by this shocked and accelerated CSM
disappear a few days later because they are engulfed by and hidden
inside the SN photosphere, and not because they are accelerated to 
the same speed as the SN ejecta or because the ionized CSM
recombines.  Engulfing the CSM interaction region rather than
accelerating to the same speed as the ejecta requires that the CSM is
asymmetric, as noted above, and probably indicates that we are observing the SN from some mid-latitude
direction that is offset from the equatorial plane.  This, in turn, is also
consistent with the lack of any narrow P Cygni features in any of the
high resolution PEPSI spectra.  This lack of narrow P Cygni absorption
suggests asymmetry in the CSM because the slow CSM is not seen in
absorption along our line of sight to the SN continuum photosphere,
even though it is clearly seen in emission.  This is only possible if
the CSM is not spherical.

Although we argue that the CSM is asymmetric in the case of SN~2023ixf
based on the observational behavior of the narrow and
intermediate-width lines, this may not necessarily be the case for
all SNe II with fleeting CSM interaction.  It would be interesting to
determine what fraction of these events show evidence of aspherical
mass ejection shortly before explosion, and how many seem consistent
with spherical CSM.  This may help elucidate what role binary
interaction may play in ejecting the mass, or shaping the mass ejected
by some other mechanism.

\subsection{Noteworthy things we do not see}

Here we briefly comment on a few things that we do {\it not} see in the PEPSI
spectra of SN~2013ixf, but which have been seen in early spectra of
some other SNe II with fleeting signs of CSM interaction.  These may
help us to understand how SN~2023ixf fits into the observed diversity
of this phenomenon.

1.  Except for very weak and narrow components on day 2.6, we do not 
detect lines like the C~{\sc iii} and N~{\sc iii} blend
in the ``WR bump'' just to the blue side of He~{\sc ii} $\lambda$4686,
or strong emission from narrow He~{\sc i} that has been seen in other
early spectra of SN~1998S, PTF11iqb, 2013cu, and others
\citep{shivvers15,smith11iqb,galyam14}.

2.  We do not detect a broad emission feature near $\lambda4600$ at any epoch. This broadened feature has been seen in several SNe~II-P with fleeting CSM interaction signatures \citep{quimby07,bullivant18,hoss18,andrews19,soumagnac20,pearson23,hoss22}, and 
is often attributed to broad blueshifted He~{\sc ii}
$\lambda$4686 emission from fast (5,000-10,000 km s$^{-1}$) SN ejecta crossing the reverse shock, or a blend of He~{\sc ii}
$\lambda$4686 with several other ionized features in the region.  On the other hand, very broad features may be de-emphasized by the continuum normalization in these cross-dispersed spectra if a faint broad feature crosses more than one echelle order, so low-resolution spectra may be more informative about this broad He~{\sc ii} feature.

3.  Importantly, as noted above we see no evidence for narrow P Cygni
(or any) narrow absorption from unshocked CSM.  While narrow absorption features can easily be lost in low-resolution spectra when they are seen next to a strong narrow emission feature, they are easily detected in echelle spectra \citep{smith20}.  The absence of this absorption in SN~2023ixf may indicate that its slow and dense CSM is not seen along our
line of sight to the continuum photosphere, requiring that the CSM
has a nonspherical geometry.  This would allow it to be seen in
emission out of our line of sight, but not in absorption.

\section{Summary and Conclusions}

We present a series of high-resolution echelle spectra of the recent SN~2023ixf in M101.  These provide an unprecedented record of the high-resolution emission-line evolution in a SN~II with early signs of CSM interaction with an almost nightly cadence.  These spectra reveal rapid evolution in the strength and profile shape of narrow and intermediate-width emission lines associated with CSM interaction.  Here is a summary of the main observational results:

1.  As in other SN of this class, we detect strong narrow and intermediate-width emission from H$\alpha$ and high-ionization lines such as He~{\sc ii}, C~{\sc iv}, and N~{\sc iv}.  Unlike several other SNe of this class, however, SN~2023ixf does not show strong emission from lower-ionization species like He~{\sc i} or C~{\sc iii}/N~{\sc iii}, which can be very strong in these objects.  These lines are seen, but they are very weak and limited in time during our observational window.

2.  All narrow line components fade quickly in 1-2 days from our first observation (which is a time period of 2-4 days post-explosion), and intermediate-width components of high-ionization lines linger for another 1-2 days before fading from the spectrum.  

3.  All narrow emission components show a pronounced blueshift in the earliest epochs.  The blueshift is understood as resulting from a combination of light travel time effects and occultation of the far side of the CSM by the photosphere.  However, the amount of blueshift and the width of the narrow component depends on both time and on the ionization level of the line.  Higher ionization lines are broader and more blueshifted than lower ionization in our first epoch, and this difference with ionization level diminishes after a day with all lines showing roughly the same width and blueshift as the higher ionization species.  This requires either acceleration of the innermost dense CSM, or asymmetric CSM.

4.  The H$\alpha$ wings in our first epoch are consistent with electron scattering wings (i.e. they are well fit by a symmetric Lorentzian shape, with a centroid that has a similar blueshift as the narrow component).  However, this changes after 1-2 days.  H$\alpha$ and He~{\sc i} lines develop intermediate-width P Cygni absorption, requiring that the broadening of these lines after the first day or two is tracing kinematic expansion and is not due only to electron scattering.  The P Cygni absorption indicates expansion speeds of 700-1,300 km s$^{-1}$, tracing CSM that has been swept up in to the post-shock shell.

5.  As the intermediate-width components fade, the observed velocities do not increase.  The P Cygni absorption, in particular, remains steady at $<$1300 km s$^{-1}$.  This requires that the CSM interaction signatures are not fading because the post-shock gas is getting accelerated and incorporated into the fast SN ejecta.  Instead, the CSM interaction region is likely asymmetric and gets engulfed and hidden by the SN photosphere.

6.  Although the narrow components are easily resolved in our echelle spectra, none of our spectra show narrow P~Cyg absorption from dense pre-shock CSM along our line-of-sight to the continuum photosphere.  This may require that the CSM is asymmetric. 

7.  The width of the narrow H$\alpha$ component indicates a CSM expansion speed of about 115 km s$^{-1}$, and this is seen in our first epoch before the H$\alpha$ appears to get accelerated to the same blueshift and width as the higher ionization lines.  This expansion speed is 5-10 times faster than a normal RSG wind.

8.  The disappearance of the CSM interaction signatures after a few days suggests that the CSM is confined to a relatively compact radius of 20-30 AU (or $\la$10$^{14.7}$ cm).  This radius, combined with its observed expansion speed of 115 km s$^{-1}$, implies that the CSM was ejected roughly 1 yr before core collapse.  

Altogether, we find several clues that the confined CSM of SN~2023ixf is asymmetric.  We interpret the evolution of the line profiles as indicating that the asymmetric CSM interaction region is engulfed by the SN photosphere (e.g., \citealt{smith11iqb}).  While the narrow lines may weaken because the pre-shock gas is accelerated and incorporated into the post-shock shell, the resulting intermediate-width lines are hidden from view behind the SN photosphere, rather than the high-ionization lines fading away because the shocked shell is accelerated or because the gas recombines.  While the inferred timescale for creating SN~2023ixf's pre-SN CSM is about a year (which suggests an instability during Ne or O burning), the implied asymmetry in the CSM points to a scenario where pre-SN inflation during Ne or O burning will instigate binary interaction that ejects mass into a disk or torus.  Thus, CSM interaction may continue to occur as SN ejecta hit the engulfed CSM, but it may be hidden from our view.  Depending on the mass of the CSM, CSM interaction signatures may reappear after SN~2023ixf drops from its pleateau when the recombination photosphere recedes again.

\vspace{5mm}

\newpage

\begin{acknowledgments}

We thank an anonymous reviewer for comments that improved the
paper. For help with obtaining and reducing the spectra, we thank LBT
staff members including Alex Becker, Jennifer Power, and director Joe
Shields.  Some of these LBT/PEPSI spectra were obtained as part of a
pre-approved program called AZTEC (Arizona Transient Exploration and
Characterization), but some resulted from Director's Discretionary
Time and Engineering time, which allowed a nearly nightly cadence
during a critical time period. The LBT is an international
collaboration among institutions in the United States, Italy and
Germany. LBT Corporation partners are: The University of Arizona on
behalf of the Arizona university system; Istituto Nazionale di
Astrofisica, Italy; LBT Beteiligungsgesellschaft, Germany,
representing the Max-Planck Society, the Astrophysical Institute
Potsdam, and Heidelberg University; The Ohio State University, and The
Research Corporation, on behalf of The University of Notre Dame,
University of Minnesota and University of Virginia.

We respectfully acknowledge the University of Arizona is on the land
and territories of Indigenous peoples. Today, Arizona is home to 22
federally recognized tribes, with Tucson being home to the O’odham and
the Yaqui. Committed to diversity and inclusion, the University
strives to build sustainable relationships with sovereign Native
Nations and Indigenous communities through education offerings,
partnerships, and community service.
  
Time domain research by D.J.S.\ and team is supported by NSF grants
AST-1821987, 1813466, 1908972, \& 2108032, and by the Heising-Simons
Foundation under grant \#20201864.
This publication was made possible through the support of an LSSTC
Catalyst Fellowship to K.A.B., funded through Grant 62192 from the
John Templeton Foundation to LSST Corporation. The opinions expressed
in this publication are those of the authors and do not necessarily
reflect the views of LSSTC or the John Templeton Foundation.

\end{acknowledgments}

\vspace{5mm}

\facilities{LBT:PEPSI}




\bibliographystyle{aasjournal}
\bibliography{refs}


\end{document}